\documentclass[aps,preprint,a4paper,prc,showpacs,showkeys,
nofootinbib]{revtex4-1}
\usepackage{graphicx}
\usepackage{bm} 
\usepackage{amssymb,amsmath,amsfonts}

\begin{document}

\title{Eta-nuclear interaction: optical {\it vs.} coupled channels}
\author{J.A. Niskanen}

\affiliation{Helsinki Institute of Physics, PO Box 64, 
FIN-00014 University of Helsinki, Finland }
\email{jouni.niskanen@helsinki.fi}
\date{\today}

\begin{abstract}
The existence of possible $\eta$-nuclear bound states
is closely related to the corresponding scattering lengths.
While the sign of its real part may indicate a bound state,
a large (always positive) imaginary part can prevent
such a state. Most theoretical calculations for {\em e.g.}
$^3$He predict quite sizable imaginary parts with no
bound state. It is shown that a generalization of 
the conventional
phenomenological optical model potential to coupled 
channels, based otherwise on the same assumptions but 
treating the pionic channel explicitly, can yield much
smaller inelasticity still starting with elementary $\eta N$
interactions giving the same $\eta N$ scattering lengths.
As representative examples this decrease is argued by
model calculations in the case of $\eta\,^3$He and
$\eta\,^{12}$C.
\end{abstract}


\maketitle

\section{Introduction}
Since the realization by Bhalerao and Liu 
\cite{bhalerao} that the $\eta N$ interaction
is relatively attractive the next step was an
anticipation of possible $\eta$-nuclear (quasi)bound
states \cite{haider,haider2}. In spite of intense searches,
so far no unambiguous experimental evidence has been brought
up to support these expectations\footnote{Ref. 
\protect\citep{mg25}
reports a possible observation in $^{25}$Mg.}. 
Also theoretical predictions are mixed, varying from
bound states for nuclei only heavier than carbon to claims
of binding for $^4$He or even $^3$He \footnote{For an 
extensive recent review see Ref. {\protect\cite{machner}}}. 

The existence of bound states is closely related to
scattering, in particular to the low energy 
expansion by the scattering length and effective range
\begin{equation}
q \cot\delta = \frac{1}{a}+ \frac{r_0}{2}q^2\, ,
\label{efran}
\end{equation}
where, with the convention normal in meson physics,
a positive $\Re a$ indicates moderate attraction, while a 
negative value means repulsion or a bound state.
Unfortunately, this relation is predictive only in theory,
since experimentally the cross section in scattering or in
production final state interactions (FSI) cannot distinguish
the sign of the real part.

However, as pointed out earlier by Haider and Liu, actually the 
condition for complex potentials is more restrictive and also
$|a_{\rm R}| > a_{\rm I} $ should be valid \cite{haider3}. 
By unitarity,
the imaginary part $a_{\rm I}$ is always positive. In Ref.
\cite{Sibirtsev} this condition was pushed to the next order
in $r_0/a$ with the condition
\begin{equation}
\Re [a^3(a^* - r_{0}^*)] > 0 \; ,
\label{condition}
\end{equation}
 which reduces to the former one,
if $r_0=0$. From these conditions (albeit with the bold
assumption that $|a|\gg |r_0|$) one can see that also the
imaginary part of the scattering length has an essential
role even for the very existence of bound states, not to say
anything about their width. 
 For this reason
a detailed study and
understanding of also the imaginary part of the $\eta$-nuclear
scattering length is relevant.
In fact, a very strong correlation
between $a$ and the binding properties has been seen
for nuclei ranging from helium to magnesium
in Refs. \cite{Sibirtsev2,carbon,cracow,hailiu} giving constraints 
for the latter in the region of $(a_{\rm R},\, a_{\rm I})$ plane
where bound states could exist. This means that the FSI data can
yield information on the potential bound states only on the
condition that they exist - anyhow a possible starting point 
to make meaningful guesses 
in searches for binding observables from
scattering data.

Theoretical calculations 
for the low energy parameters to compare with experimental
FSI effects are also very varied even for the lightest
real nucleus $^3$He studied most intensively (for a
review see {\it e.g.} Ref. \cite{Sibirtsev}).
In addition to the wide variation of the predicted real part
in the case of $^3$He another problem is the predicted
imaginary part, which is often large. This is a problem 
for two reasons. Firstly, obviously the bound state 
could be too broad for observation. Secondly, even if 
$\Re a$ were negative, the above condition (\ref{condition})
 for the existence of a bound state 
may not be satisfied with a large imaginary
part. Therefore, the large predicted imaginary parts 
are a bad prospect for finding bound $\eta$-nuclear states.
However, there are indications about unexpectedly small
imaginary parts from the meta-analysis \cite{Sibirtsev}
on $^3$He and later experiments and analyses
of the $p+d\rightarrow \eta+^3$He reaction
\cite{Smyrski07,Mersmann07} and the $\eta^4$He final state 
 studied in $d+d$ interactions making use of unpolarized
beams \cite{Wronska05} as well as
polarized beams \cite{Budzanowski09b}.

In Ref. \cite{Sibirtsev} a reanalysis of existing data on
the $\eta \, ^3$He system was presented. These data stem from the
reaction $pd \rightarrow \eta\, ^3{\rm He}$ and the extraction
of the scattering length was based on the standard low energy
expression of the final state interaction
\begin{equation}
|f|^2=\frac{|f_p|^2}{1{+}a_{\rm I} q {+}|a|^2 q^2}\, ,
\end{equation}
where the original production amplitude $f_p$ is assumed to
be very short ranged and essentially momentum independent.
The global fit to then available data gave the result
$a = \pm 4.3 \pm 0.3 + i\, (0.5 \pm 0.5)$ fm.
It should still be stressed that this analysis cannot determine
the sign of the real part, which only appears in the second power.
Further, this result is fully consistent with a coupled channel
$K$-matrix analysis of Ref. \cite{Green} yielding
$a= 4.24\pm 0.29 + i\, (0.72\pm 0.81)$ fm. In both cases
the imaginary part is smaller than most theoretical predictions.

These values may be contrasted with the seemingly contradictory
results of two different more recent high-precision experiments at COSY
$a = \pm 2.9 \pm 2.7 + i\, (3.2\pm 1.8)$ fm (COSY-11 \cite{Smyrski07})
and $a = \pm 10.7 \pm 0.9 + i\, (1.5\pm 2.8)$ fm (ANKE 
\cite{Mersmann07}). In analyses the latter group also considers 
the smearing over the beam energy profile leading to the larger
real part. Further, the latter data allow also the extraction of the effective range
$r_0 = 1.9\pm 0.1 + i\, (2.1\pm 0.3)$ fm giving a better fit
than without this term  \cite{HMcracow}. 
The theoretical necessity of this term has been stressed
as well in Ref. \cite{carbon} as above in Eq. \ref{condition}.
Further support for a small imaginary part may be derived from 
an overall result for $\eta^4$He scattering length
$a = \pm 3.1 \pm 0.5 + i\, (0\pm 0.5)$ fm \cite{HMcracow}.
It is very interesting and suggestive also to note that, if the
real part for $^4$He is really smaller in magnitude than
for $^3$He, the behaviour indicates binding for $^4$He
(without any conclusion for $^3$He). The reason is that the 
heavier nucleus is probably more attractive and, in the
nonbinding situation, its scattering length should be larger.
If the binding threshold is passed, there is no constraint 
on the magnitude any more.

The aim of the present paper is to investigate possible
justification for the smallness of $\Im a$. First for
the $\eta N$ interaction the standard
static optical potential model is replaced by a coupled
channels model with an assumed explicit coupling of the
$\eta N$ system to the pion-nucleon system in a
totally phenomenological way but giving the same 
elementary $\eta N$
scattering lengths\footnote{The $\eta N$ is supposed to 
be strongly coupled to the $N^*(1535)$ baryon resonance 
leading to $\pi N$. This would give energy dependence over 
a wider range. However, this work is not concerned about this
microscopic mechanism, but just a direct coupling to $\pi N$
is assumed. The energy dependence due to the $N^*$ might
influence the effective range term.}. 
Normally the nuclear density profile
is used to spread the $\eta  N$ interaction over the nucleus
leading to single channel optical potentials, essentially 
neglecting the effect of nucleon correlations and other
nuclear ``granularity'' as well as excitations.
Now, in Sec. \ref{sec:model} this averaging approach is  
generalized to a sort of a two channel optical model.
While the limit of a complex optical potential could,
 in principle, be total absorption (``black sphere''), 
in the case of explicit two channels there
is a feedback effect. With the stronger nuclear interaction
this could make a difference by an earlier saturation of
the absorption, even though for scattering 
from a single nucleon the zero energy results would be the 
same. Another factor could be the longer range of a nucleus 
{\it vs.} the large wave number of the pionic channel.
In Sec. \ref{sec:results} this turns out to be the most 
important effect. 
With a single channel optical potential the size of the 
nucleus does not play a particularly important role in
zero-energy scattering, but for the coupled high momentum
pion channel the soft form factor decreases the 
transition probability dramatically.

\section{\label{sec:model}Coupled optical model}
In line with the simple optical approach \cite{wilkin}
the $\eta N$ and $\eta$-nuclear potential can be expressed as
\begin{equation}
V_{\rm opt} = -4\pi (V_{\rm R} + i V_{\rm I})\, \rho(r)\, 
\hbar^2/ (2\mu_{\eta N})\, ,
\label{potential}
\end{equation}
with $\mu_{\eta N}$ the reduced mass of the $\eta N$ system
and $\rho$ the nuclear density ($V_{\rm R}$ and $V_{\rm I}$ 
in fm). In Ref. \cite{wilkin} the strength parameters are 
taken to be the complex scattering length.

This form is used to produce the $\eta N$
scattering length. Unfortunately, this quantity may not be 
very well known with values for its real part varying roughly 
between
about 0.25 fm ( {\it e.g.} chiral models \cite{chiral}) and
about 1 fm ({\it e.g.} K matrix methods \cite{kmatrix}) 
and the imaginary part between 0.2 fm and 0.4 fm. An 
up-to-date listing can be found in Ref. \cite{machner}.
However, most of the analyses for $\eta N$ scattering length
 yield the magnitude of
the imaginary part roughly equal to one half of the real part.
K matrix methods tend to give lower ratios down to a quarter
and chiral models higher, but in this calculation, just to
compare the effect in nuclei for {\it scattering length
equivalent elementary
interactions}, the ratio is kept as one half. So $a_{\rm I} = 
0.5\, a_{\rm R}$ and the strengths $V_{\rm R}$ and $V_{\rm I}$ 
will be varied so that $a_{\rm R}$ covers the interval 
0.2 -- 1 fm.  In the case of
the elementary interaction the range is obviously short,
rather dictated by the size of the hadrons. In this case 
the density profile is taken simply as a normalized Gaussian
\begin{equation}
\rho(r)= A\, \exp [-(r/b)^2]/(\sqrt{\pi}\, b)^3\, ,
\label{etanprof}
\end{equation}
where $b$ is the range parametre and $A=1$.
 
In the simplest static optical potential the strength 
parameters $V_{\rm R}$ and $V_{\rm I}$ are sometimes
 taken to be the components of the zero energy elementary 
amplitude ({\it i.e.} the scattering length $a_{\eta N}$
as in Ref. \cite{wilkin}).
This may be thought of as spreading over the nuclear size
the scattering strength from single nucleons. An
implicit background assumption could be a
density profile of Dirac's $\delta$-functions,
point-like sources. However, it was numerically found that
this assumption cannot be used for a potential approach.
It was impossible to make the range $b$ arbitrarily small
in the Schr\"odinger equation for any constant strength
$V_{\rm R}$. This is due to the fact that qualitatively
a condition for bound states (and the associated singularities 
in the scattering length) with varying potential strength and 
range is that the well-depth times the squared
range should be 
larger than some constant. (In the case of a square well
$\pi^2\hbar^2 /(8\mu)$.) 
However, making the range smaller, but at the same time
increasing the normalization constant as required by the 
$\delta$-function limit, causes the well 
effectively to deepen inversely
to the cube of the range, as can be seen from eq. 
\ref{etanprof} and the above binding condition will be met. 
(For a real square well the resulting binding condition would
be $R < 12\, V_{\rm R} / \pi^2$ and presently for the Gaussian
$R < 0.84\, V_{\rm R}$.) With still decreasing range
more bound states and scattering length singularities would
appear and pass.
The importance of the distortions in the context of 
short-ranged strong interactions has been discussed 
in {\it e.g.} Ref. \cite{peierls} in the case of repulsive
interactions, but the effect for attraction is even more
drastic and achieving a $\delta$-function meaningfully seems
impossible.

For the coupled channels interaction the model to be used
is similar, but in eq. (\ref{potential}) the strengths
will be matrices. In that case
$V_{\rm R}$ is replaced by a diagonal $2\times 2$ matrix 
and $V_{\rm I}$ effectively by an off-diagonal 
$\eta N\leftrightarrow \pi N$ transition matrix. Let's
denote its strength as $V_{\rm C}$ for coupling.

Since both the real and imaginary parts of the $\eta N$
scattering lengths can be described by just two interactions,
for the present discussion only the interactions primary 
to the $\eta N$ sector are considered, 
{\it i.e.} the diagonal $\pi N$ potential is neglected.
This is purely a practical choice to avoid inessential
complications in the present intent to point out 
in principle the
important difference between single- and two-channel
optical potentials. The omission, of course, influences
$\pi N$ scattering, which, anyway, would give
mainly an overall elastic extra phase to the transition
matrix. Being attractive, added to the
$\eta\pi$ mass difference in 
Eq. \ref{pinequ} the $\pi N$ potential would increase 
locally the already large $\pi N$ wave number, 
further decreasing the transition matrix as will be 
discussed later, and corroborating the case even more.
Also its secondary effect to the inelasticity is anyway 
to some extent assimilated by the phenomenological 
variation of $V_{\rm I}$ and $V_{\rm C}$. It may be noted
in passing that in $\pi N$ scattering the $\eta N$ threshold
in the energy region close to the $\eta\pi$ mass difference
(and the $N^*(1535)$)
causes strong attractive peaking taken automatically into
account by coupled channels even without an explicit
diagonal potential in this channel.

The single channel Schr\"odinger equation is perfectly 
standard
\begin{equation}
\frac{{\bf p}^2}{2\mu}\, \psi + V_\eta\,\psi = T\,\psi
\end{equation}
with $V_\eta$ the complex $\eta$ potential (\ref{potential}), 
$T$ its
kinetic energy and $\mu$ the relevant reduced mass. 

In the case of the coupled model also the pion wave 
function appears into the radial equation ($s$-wave)
\begin{equation}
\frac{d^2u_\eta}{dr^2} - \frac{2\mu}{\hbar^2}\, V_{\eta}(r)
\, u_\eta(r) - \frac{2\mu}{\hbar^2}\, V_{\eta\pi}(r)
\, u_\pi(r) = - \frac{2\mu}{\hbar^2}\, T
\, u_\eta(r)\, 
\end{equation}
with $V_{\eta\pi}$ the transition potential (of strength 
$V_{\rm C}$)  and $\mu$ the relevant reduced mass.
The light pion with the total energy equal to the $\eta$
mass cannot be handled by the same equation but the
relativistic version (Klein-Gordon equation) is more
relevant. Here the local momentum is represented by the
modified Einstein relation $p^2 = (E_{\rm tot}-V)^2
-m_\pi^2 c^4$ leading in the lowest order in $T/m_\eta c^2$
and $V/m_\eta c^2$ to
\begin{equation}
\frac{d^2u_\pi}{dr^2} + \frac{(m_\eta^2 - m_\pi^2) c^4}
{(\hbar c)^2}\, u_\pi(r)
- \frac{2m_\eta}{\hbar^2}\, V_{\pi}(r)
\, u_\pi(r) - \frac{2\mu}{\hbar^2}\, V_{\eta\pi}(r)
\, u_\eta(r) = - \frac{2m_\eta}{\hbar^2}\, T
\, u_\pi(r)\, . \label{pinequ}
\end{equation}
In the transition term of the pionic channel the $\eta$ mass 
has been replaced by the reduced mass of the first channel
for hermiticity. It should be noted that in the
calculation of the elementary scattering this strength is
anyway a freely adjustable parameter and for nuclear
scattering the difference is not large. As already 
mentioned the diagonal $V_\pi$ term will be neglected.
From the asymptotics one gets for the free pion momentum
\begin{equation}
q_\pi =  \sqrt{ \frac{(m_\eta^2 - m_\pi^2) c^4 
+2m_\eta c^2 T}{\hbar^2 c^2} }
\label{pimom}
\end{equation}
to be used in the asymptotic boundary conditions.


The procedure is then, after finding the numerical correspondence
between the elementary $(V_{\rm R},V_{\rm I})$ or
$(V_{\rm R},V_{\rm C})$
giving the same $(a_{\rm R},a_{\rm I})$
for both models, to use the strengths thus obtained 
for a given $a$ to calculate nuclear scattering with 
more extensive nuclear density profiles (normalized to $A$). 
This means simply summing the potentials of individual 
nucleons and phenomenologically smoothing the
resultant total potential to the nucleon density in the 
philosophy of single channel optical potentials.  This is 
not, however, in detail quite the same as in 
{\it e.g.} Ref. \cite{wilkin}, where the scattering lengths
were used as the strengths instead of $\eta N$ potentials. 
Rather this is the ``direct'' 
interaction part for the optical potential \cite{joachain}.

The simple smooth averaging process necessarily
suppresses quite a lot of microscopic structure mechanisms.
Still in this respect the present treatment is no worse
or better than the conventional optical potential model
of Ref. \cite{wilkin}. The point of this calculation is
to show that the explicit inclusion of the low-threshold 
$\pi$-nuclear channel with large momentum is essentially
different from the treatment of the inelasticity  by the
imaginary potential component in the $\eta$-nucleus
channel, leading to its strong decrease. This fundamental
difference seems to have been largely overlooked so far in 
$\eta$-nuclear discussions.

Now, in  the coupled case with the enhanced nuclear
transition potential the feedback effect from pions should be 
larger and consequently the inelasticity could be smaller than
in the direct single-channel optical model. In fact, quantum 
mechanically
the strong coupling limit should be to share the probability
of $\eta$'s and pions equally instead of formally total absorption
in the optical model limit. However, one should keep in mind 
that strong absorption is nonlinear 
in $V_{\rm I}$ and quite complex. The
imaginary potential eats the wave function off and acts like
repulsion causing correlations, which tend to saturate 
possible inelasticity as seen {\it e.g.} in Ref. \cite{carbon}
for $\eta$-nuclei and in even stronger annihilation of
antinucleons \cite{anti}. Another effect to corroborate
the expectation of weaker absorption in the coupled model 
is the large wave number in the pionic
channel (minimum 2.7 fm$^{-1}$) forcing by oscillations the
relevant transition matrix to decrease for smooth long ranged
nuclear potentials. In fact, as seen later numerically
this effect can be a decrease of inelasticity by 
orders of magnitude and in the momentum representation would
be considered to be a form factor effect. If the $\pi N$
diagonal potential is included, its effect would be to 
change the wave number somewhat, with attraction effectively
increasing its value.

\section{\label{sec:results}Results}
As discussed in the previous section, trying to make the
$\eta N$ interaction range infinitesimally small is 
impractical or even impossible. So as the finite extension 
$b$ in eq. (\ref{etanprof}) 0.3 fm is adopted at first.
For this choice the strength parameter $V_{\rm R}$ varies
roughly between 0.14 -- 0.28 fm in the case of the single
channel optical potential and between 0.10 -- 0.25 fm for 
coupled channels to produce the values of $a_{\rm R}$
in the interval 0.2 -- 1.0 fm.
The imaginary (or coupling) strength 
varies between 0.04 -- 0.03 fm and 0.14 -- 0.12 fm,
respectively. It may be noted that
without absorption and the subsequent effective
repulsion the upper values would be close to a binding
strength for such a short range, as discussed earlier. 
The dependence on the elementary range will be studied later.

\begin{figure}[tb]
\includegraphics[width=\columnwidth]{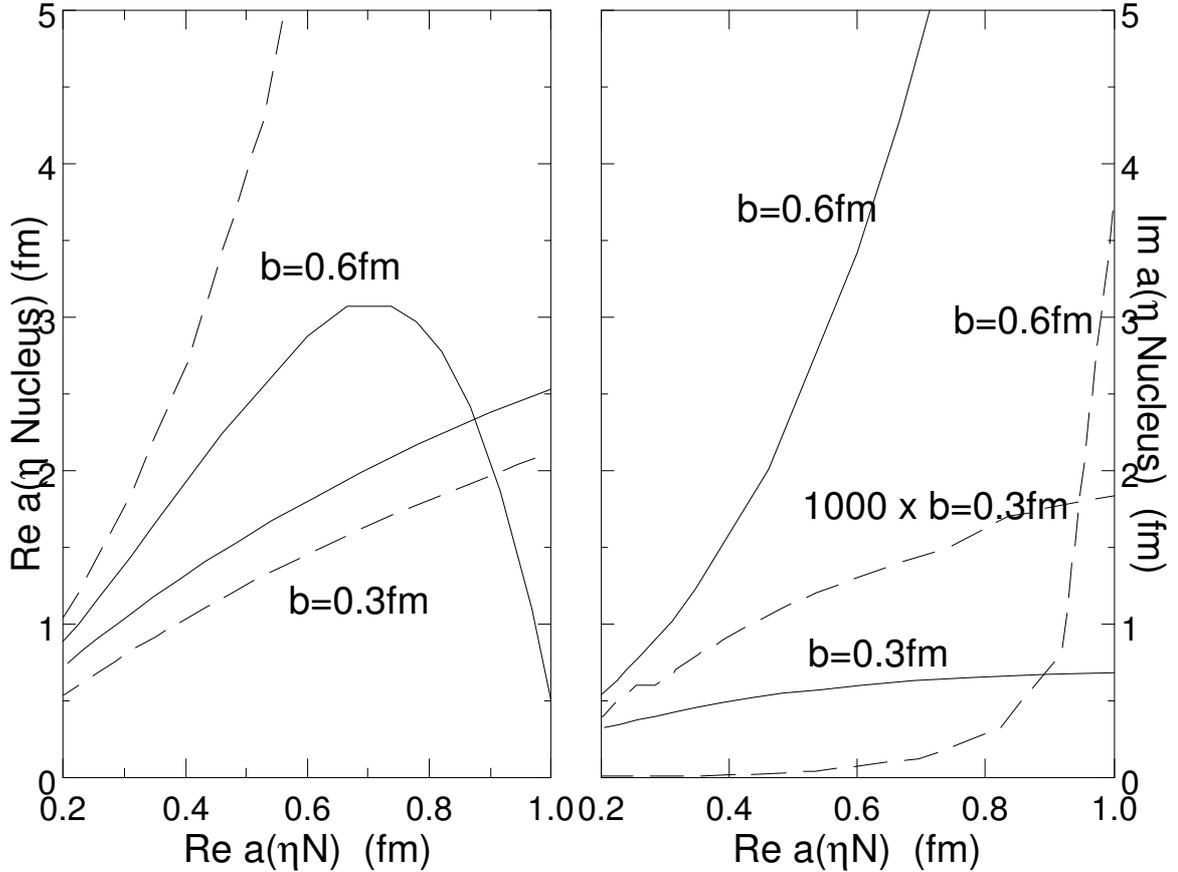}
\caption{Nuclear scattering length $a(\eta ^3$He)
 as a function of the
elementary $\eta N$ scattering length calculated for 
potentials yielding the same elementary lengths
(with the constraint $a_{\rm I} = a_{\rm R} / 2$).
Solid: single channel optical model, dashed: coupled
channels. The two values of the $\eta N$ range parameter 
$b$ are indicated. \label{nuclength}
}

\end{figure}

As the most relevant and most investigated nucleus $^3$He 
is used as an example with a Gaussian
profile as given in eq. (\ref{etanprof}) but using the range parameter 
$b(\eta^3{\rm He}) = \sqrt{2/3}\, r_{\rm rms} = 1.55$ fm with
the root-mean-square radius of 1.9 fm and the normalization
to $A=3$ \cite{wilkin}.
The results for the real and imaginary parts of the 
nuclear scattering length is given in Fig. \ref{nuclength}
as functions of the elementary $a(\eta N)$.
While the real parts differ only moderately, in the 
imaginary parts there is a dramatic difference of more than 
two orders of magnitude (solid {\it vs.} dashed). This
decrease is interesting, even though the real parts
in this case remain positive, {\em i.e.} nonbinding.

To investigate the origin of the drastic drop in 
the imaginary part of the nuclear scattering length the 
calculation of the effect is divided into a study of
two possible mechanisms indicated previously.
First the effect of changing the strength only
 is shown, then a change only in the range. The varying 
is performed superficially by a multiplicative
factor acting on both model strengths
 $(V_{\rm R},V_{\rm I})$ or $(V_{\rm R},V_{\rm C})$ giving 
originally the same single
elementary scattering length a($\eta N)$ = (0.55 + 0.27 i) fm.
The optical model strength for this is $(V_{\rm R},V_{\rm I})
= (0.23,0.0393)$ fm and coupling $(V_{\rm R},V_{\rm C}) =
(0.20,0.14)$ fm.

\begin{figure}[tb]
\includegraphics[width=\columnwidth]{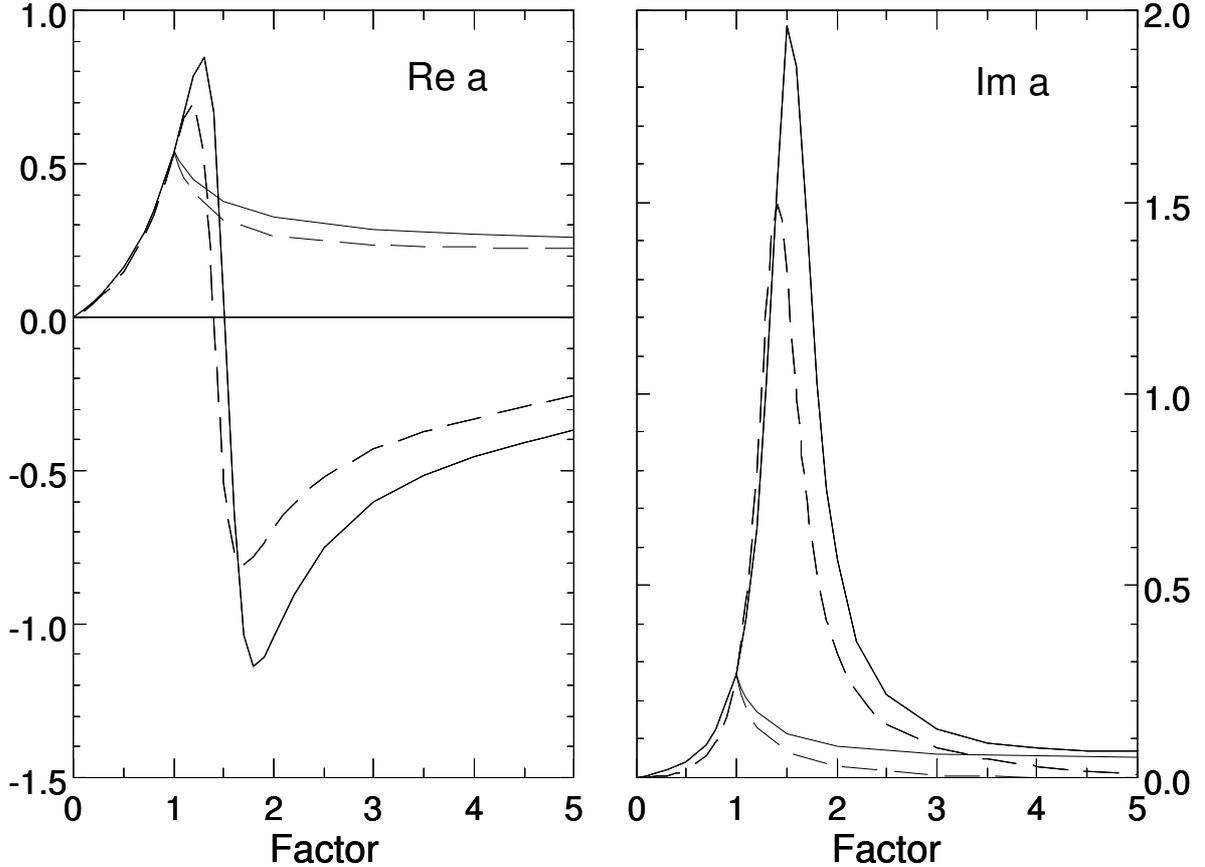}
\caption{Superficial
$\eta N$ scattering lengths for an optical
potential (solid) {\it vs.} coupled model (dashed) 
as a function of strength and range. 
The elementary $\eta N$ models 
yielding the same scattering length (0.55 + 0.27 i) fm
have been modified varying either the strength or range by
a multiplicative factor between 0 and 5.
The curves starting from 0 vary the strength, whereas those
starting discontinuously
from factor 1 correspond to varying the range. \label{facfig}
}

\end{figure}

The results are shown in Fig. \ref{facfig}.
It can be seen that as the strength $(V_{\rm R},V_{\rm I})$
or $(V_{\rm R},V_{\rm C})$ doubles, both the 
real and imaginary parts experience qualitatively strong
variation. The real part on the left shows first strong 
attraction changing quickly into apparent
repulsion with sharp maxima in both models.
This behaviour could also reflect a complex bound state,
as this limit for the very short ranged interaction is close.
(Of course, the elementary interaction does not support
this.)
The imaginary part on the right-hand panel
goes through a sharp peak in the same interval. 
Apparently the strong absorption causes the change into 
effective repulsion. Also the maximum expresses a
saturation of absorption. Although these strength
varying curves are qualitatively similar for both 
models, it is noteworthy that for a stronger
interaction the coupled channel result has a much smaller
imaginary part. If the strength of several nucleons were
concentrated within the elementary range, the coupled 
channels would give less absorption according to this
calculation.

However, real nuclei have an extension much larger than
0.3 fm. The less dramatically behaving curves starting
from ``factor''  $=1$ describe corresponding changes due to
multiplying the range by this factor. (It does not make
sense to study smaller ranges.) 
The behaviour of the real part is now smooth and similar
in both models. Also the imaginary part does not look
particularly spectacular, but it is important to note 
that in the case of
coupled channels the vanishing with increasing range
is very much faster than for the optical potential, which, 
combined with the strength variation,
could account for the unexpectedly small result for
$^3$He. It may further be noted that with the 
Gaussian distribution the upper limit in the figure 
would actually closely correspond to the nuclear $^3$He 
distribution with ``factor'' = 5.17. For this value of
``factor'' the imaginary part $a_{\rm I}$ in the
coupled channels model is already vanishingly small.

In the case of real nuclei both effects play their
roles. The nuclear size increases with $A$ as well  as the
strength does. The latter, however, is moderated by the
volume (and hence by the range) and eventually saturates.
The influence of the size can be thought as a form factor 
effect
for the case of coherent inelasticity with the nucleus
remaining intact. This is actually an inherent assumption
in the simplistic optical model with the potential 
described as being proportional to the density, but the form 
factor effect really hits only in the explicit inelastic
pion channel with a large wave number, not on the low 
energy $\eta$ meson. Such a strong suppressing effect for
pionic inelasticity was already suggested in Ref.
\cite{smallima} as a ratio of the nuclear and elementary
form factors.

It is time to discuss the model dependence. The basic 
interaction has been taken so far very short-ranged to
simulate a $\delta$-function potential. As discussed,
this has problems. Also the drastic behaviour in Fig.
\ref{facfig} might be an artifact due to this. 
Therefore, next the
same calculation is repeated with the range $b = 0.6$ fm,
which is certainly reasonably large. The corresponding 
results are also shown in Fig. \ref{nuclength}. Now the size 
of both the real and imaginary parts is much larger, since -
with the longer range as discussed before in sec. \ref{sec:model}
- the elementary strength also must be larger and
this factor is conveyed to the nuclear potential (whose
range is not changed). Although this change with the range
is even qualitative, still the imaginary part remains 
much smaller in the coupled case than for the optical 
one over the whole range of realistic
values of the elementary scattering length. The smallness
of the imaginary part is further emphasized by the larger
size of the real part for the coupled-channel calculation.
The rapid rise of $\Im a(\eta^3{\rm He})$ for $\Re a(\eta N)
> 0.9$ fm combined with a large maximum in the real part
is associated with the onset of a narrow bound state for
$\Re a(\eta N) \approx 1.1$ fm. Qualitatively the 
behaviour of $a(\eta^3{\rm He})$ would be similar to Fig.
\ref{facfig} though with numerically much larger values,
if this threshold is passed.

Next the numerical (in)significance of the neglect of
the diagonal potential in the $\pi$-nucleus channel was
checked by a short calculation. The addition into this
channel of the same rather strong attraction as for 
$\eta$ increased the real part of the scattering length
slightly, a few per cent in the lower half of the
$a_{\rm R}(\eta N)$ range and up to 10-20\% in the upper 
half.
The already very small imaginary part was roughly halved.
This may be regarded as confirmation of the expectation
presented in Sec. \ref{sec:model}.

\begin{figure}[tb]
\includegraphics[width=\columnwidth]{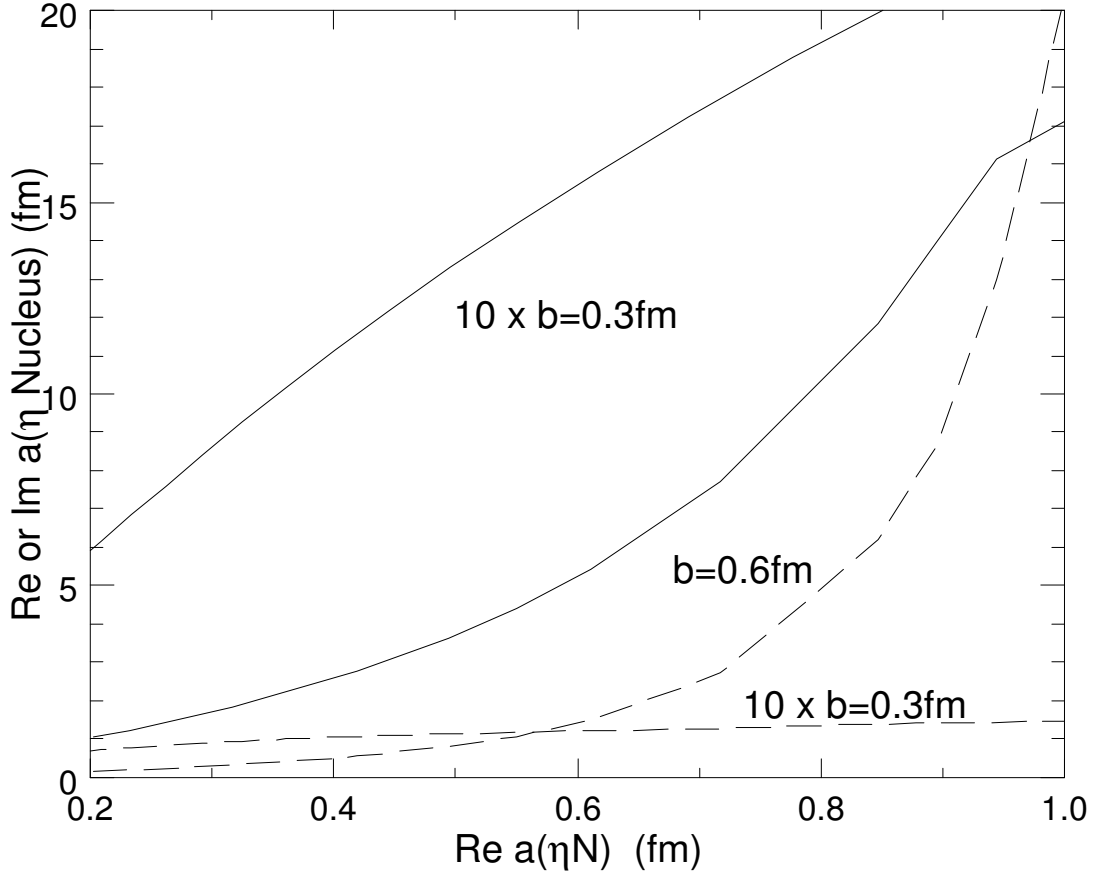}
\caption{$\eta^3$He scattering length for the coupled model  
as a function of the elementary $\eta N$ scattering length
supplemented by an optical potential to account for
 the inelasticity to two pions. 
The solid curves present the real part for two values of the
elementary range while the dashed ones are for the imaginary 
part.
\label{twopi}
}
\end{figure}

Another aspect of model dependence in the above calculation
is that pionic inelasticity is not the only one in the 
$\eta N$ system. About a quarter of inelasticity can be due
to two-pion final states to be taken into account in any
more realistic coupled optical model.
 This estimate is consistent
with the reported branching ratios of the $N^\ast(1535)$,
50\% to $\eta N$ and 13\% into $\pi\pi N$ \cite{partdata}). 
Its influence is estimated by adding to the coupled 
channels calculation also an imaginary $\eta N$ potential.
The strength of this additional potential is taken from the
conventional optical potential model yielding this fraction. 
Actually about one quarter of the earlier imaginary part
$V_I$ turns out
to be a fairly good value. Then the new complex coupled 
model becomes too absorptive (and less attractive) and its 
transition potential must be reduced to yield still 
$a_{\rm I}(\eta N) = a_{\rm R}(\eta N)/2$. The elementary 
scattering length changes only by a few percent giving
credibility to this procedure. The results of this modification
are shown in Fig. \ref{twopi} for the $\eta ^3$He scattering 
length with the two values of the elementary range
parameter $b = 0.3$ fm and 0.6 fm. As expected, because of the 
sensitivity to the optical potential for small coupling strengths
the imaginary part increases significantly, even qualitatively. 
However, for the most reasonable values of the elementary 
scattering length it still remains much smaller than for the
pure optical model results (solid curves in the right-hand
panel of Fig. \ref{nuclength}).
 Also one may note the generally increased size of the
real part in both coupled models, though the sharp peaking
is smeared quite a lot with this moderate extra absorption.

\begin{figure}[tb]
\includegraphics[width=\columnwidth]{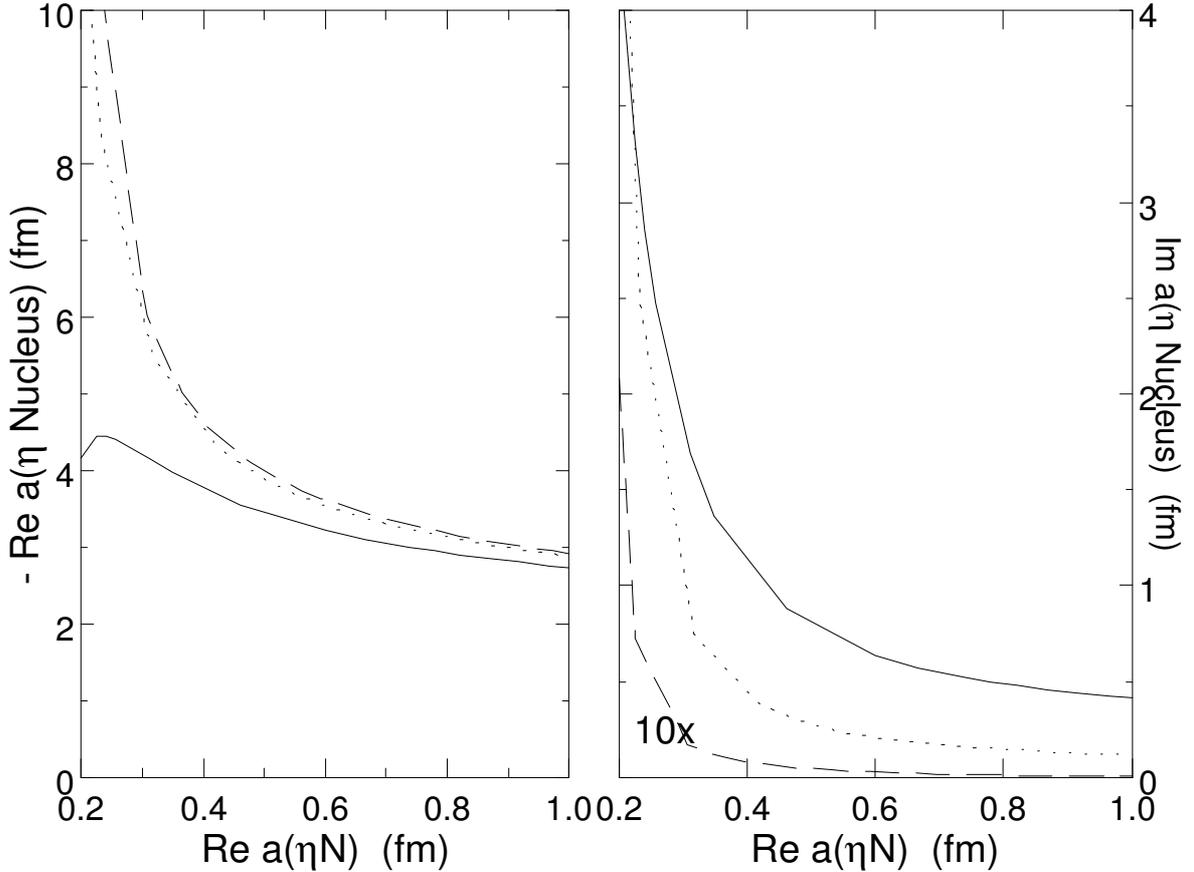}
\caption{$\eta$-nuclear scattering length on carbon 
as a function of
elementary $\eta N$ scattering length calculated for 
models yielding the same elementary lengths
(with the constraint $a_{\rm I} = a_{\rm R} / 2$).
Solid: single channel optical model, dashed: pure coupled
channels (the imaginary part multiplied by 10), 
dotted: coupled channels plus two-pion inelasticity. 
The range parameter $b = 0.6$ fm is used throughout.
Note that the real part of the amplitude is negative.
\label{carbon} }
\end{figure}

Since it has been seen that the size
of the nucleus is of paramount importance
in the coupled channel model of inelasticity, it would be 
interesting also to consider scattering from an even larger
nucleus. As a representative example let us take
 $^{12}$C, where binding is unanimously assumed. For
this the modified harmonic oscillator of Ref. \cite{atomic} may
be used as the density profile
\begin{equation}
\rho(r) = 0.17\, [1 + 1.15\,( \frac{r}{1.672\,{\rm fm}} )^2 ]
\exp{[-(r/1.672\,{\rm fm})^2]}\; {\rm fm}^{-3}
\end{equation}
with the normalization $4\pi\,\int_0^\infty \rho\, r^2\, dr = 12$.
Accordingly the optical, the pure coupled channel and the
smeared coupled channels models are applied to produce Fig.
\ref{carbon} but now only with the more realistic range
parameter $b = 0.6$ fm. In this case one may note that the real
part is negative as it should be for a binding potential.
The singular threshold is below $\Re a(\eta N) = 0.2$ fm, so 
the magnitude of both real and imaginary parts is decreasing
instead of increasing as in Fig. \ref{nuclength}.
Again the pure coupled channels model gives an extremely small
imaginary part, except very close to the binding threshold
just below the interaction strengths present in the figure. 
In that region the single channel optical model and the 
complex coupled channels become comparable.

It is noteworthy that for fairly well binding 
strong interactions both coupled channel models
give smaller imaginary parts than the single channel optical
one. This gives hope for distinguishing fairly narrow states, 
if the binding is strong enough. However, in the weaker end
of the elementary interaction ($\Re a(\eta N)\alt 0.3$ fm)
the peaking of the coupled channels result for $a(\eta A)$
(with larger magnitude of the real part) means also a smaller 
binding energy in the coupled case making the threshold final 
state interaction enhancement in $\eta$ production very sharp
and the binding energy possibly also smaller than the width.
Here the largish scattering length makes it possible to
use the low-energy expansion for the 
complex bound state energy
\begin{equation}
E = - \frac{\hbar^2}{\mu r_0^2}\, \left(1 + \frac{r_0}{a}
- \sqrt{1 + \frac{2r_0}{a}} \right)\, 
\end{equation}
to estimate these. For the most realistic model (the
combination of the optical and coupled channels) with
$\Re a(\eta N)\leq 0.25$ fm the real and imaginary parts are,
indeed, comparable, but already $\Re a(\eta N)\approx 0.3$ fm
seems feasible with $E \approx -(2 + i)$ MeV. For
strong binding, when $\Re a \approx -2\,\Re r_0$, the meaning
of this simple approximation becomes dubious even though the 
imaginary parts are small, meaning, in principle, a narrow
state. It shows the general importance of also the effective
range.

\section{Conclusion}
A coupled channels generalization of the optical potential
has been applied to
low energy $\eta$-nuclear scattering to study the effect 
of the pionic inelasticity more explicitly.  Compared with the 
simple single-channel optical model a strong 
decrease was seen in the imaginary part of the scattering
length, though the elementary $\eta N$ interactions
had been adjusted to give the same scattering lengths
and the same procedure was used to relate the $\eta$-nuclear
interaction profiles and strengths.
In Fig. \ref{facfig} this decrease was traced to
both a possible increase of the transition strength in nuclei 
{\it vs.} elementary $\eta N$ scattering (stronger 
feedback effect) and even more
importantly to the larger spatial size of nuclei. Due to 
the high pion channel momentum the softer form factor of
the sizeable nucleus decreases the 
$\eta\pi$ transition amplitude 
drastically as compared to low-energy $\eta$ inelasticity 
obtained from the single channel optical potential even though 
the elementary low-momentum $\eta N$
scattering is equivalent. This effect is strongly dependent
also on the range of the elementary interaction, {\it i.e.}
on the relation of the elementary to the nuclear form factor
as anticipated earlier \cite{smallima}, but for the range of
values normally considered reasonable for the elementary
amplitude \cite{machner} the conclusion appears valid.
This holds still after the inclusion of two-pion inelasticity
described phenomenologically by an additional optical potential
as seen in the final results of Figs. \ref{twopi} and \ref{carbon}.

Some additional nuclear contributions to $\eta$ inelasticities 
(notably absorption on nucleon pairs) were also qualitatively 
estimated in Ref. \cite{smallima} to be small, so that 
the minor imaginary parts of the nuclear scattering length 
referred to in the Introduction 
\cite{Sibirtsev,Smyrski07,Mersmann07,Wronska05,Budzanowski09b}
may have some theoretical understanding and justification.
The results also may facilitate finding $\eta$-mesic nuclei,
although the present "toy model" cannot profess to be a
full or even comprehensive calculation of such systems. 
As discussed in the context in Secs. \ref{sec:model} and
\ref{sec:results}
the seemingly important simplification of omitting the
diagonal $\pi N$ channel potential should not change 
the main conclusion of significantly reduced $\eta$-nuclear
inelasticity in this two-channel extension of the optical
model in comparison against the single-channel approach.
Further, there is no apparent reason how or why the 
extension of the optical model considered here would change
the phenomenological and numerical connection between the 
low-energy scattering parameters and $\eta$-nuclear binding
properties \cite{Sibirtsev2,carbon,cracow}.

As a cautionary note one should, however, remember that the
simple optical model potential (also with the present
extension), being proportional to the nuclear
density, does not formally take into account the change of
the nucleus and its wave function ({\it e.g.} by removal of a 
recoil nucleon), so calculations to overcome this restriction 
would be desirable. Clearly nuclear low-energy excitations
cannot be addressed by this kind of phenomenologically 
averaged optical potential approach either. Such excitations
can be important already as recoil effects in the pionic
channel. Further, as shown {\it e.g.} in Ref. \cite{gal},
the bound-state properties are also affected by subthreshold 
medium effects, which are not directly and obviously dealt
with above-threshold scattering. 
Of course, after averaging over the $\eta N$ interactions
in the nuclear environment to get the smooth optical potential
(both single-channel and coupled channels) one misses the 
close-encounter peaks, sort of nuclear granularity. As in 
nuclear interactions one might consider the average to contain
the majority leaving an uneven, perhaps perturbative residual 
interaction. This may not be a problem with the very low-energy
$\eta$'s with long wave length and coarse resolution
in the standard optical model. However, in the case of the 
high pion momenta of the coupled channels even this lowered 
roughness can be important, contributing perhaps more 
than the smooth transition potential considered here (as the
inclusion of the two-pion inelasticity did). Still,
in spite of these criticisms, if the two different optical 
models presented in this paper can give such a drastic
change of inelasticity, it is feasible that some 
amount of $\eta$-nuclear inelasticity may be reduced by
an explicit coupled equation treatment of the open 
pion-nuclear channel even in more comprehensive
calculations so that the small imaginary parts referred to 
in the Introduction would become comprehensible.

\begin{acknowledgments}
I thank J. Haidenbauer, Ch. Hanhart and H. Machner for 
useful discussions. I also acknowledge the kind hospitality of
Forschungszentrum J\"ulich.
\end{acknowledgments}

\end{document}